# Content-based Text Categorization using Wikitology

Muhammad Rafi, Sundus Hassan and Muhammad Shahid Shaikh

National University of Computer and Emerging Sciences (NU-FAST)
Karachi, Sindh, Pakistan

**Abstract**
The process of text categorization assigns labels or categories to each text document according to the semantic content of the document. The traditional approaches to text categorization used features from the text like: words, phrases, and concepts hierarchies to represent and reduce the dimensionality of the documents. Recently, researchers addressed this brittleness by incorporating background knowledge into document representation by using some external knowledge base for example WordNet, Open Project Directory (OPD) and Wikipedia. In this paper we have tried to enhance text categorization by integrating knowledge from Wikitology. Wikitology is a knowledge repository which extracts knowledge from Wikipedia in structured/unstructured forms with a warping of ontological structure. We have augmented text document by exploring Wikitology fields like: {Bag of Words, titles, redirects, entity types, categories and linked entities}. We also propose and evaluate different text representations and text enrichment technique. The classification is performed by using Support Vector Machine (SVM and we have validated this experiment on 4-fold cross-validation.

**Keywords:** *Text Categorization, Machine Learning, Wikitology, Support Vector Machine, 20- Newsgroup. Reuters-21578*

## 1. Introduction

The vast amount of text available in digital form can make it difficult to efficiently access information. One way to improve the access of information is to categorize texts. Consequently, a need of automatic organization or classification of documents was felt due to exponential growth of electronic documents. Text categorization is defined as an activity of assigning a document to one or more pre-defined categories, based on content of the document [7]. There are two types of text categorization: Single Label and Multi-Label. Single-Label or Non-Overlapping Categories is defined as a technique in which exactly one category must be assigned to each document [8]. A special case of single-label text categorization is binary text categorization in which each document must be assigned to a single category from two-possible categories. For example: Email can be categorized as Spam or Not Spam. On the other hand, Multi-Label or Overlapping Categorization is defined as a technique in which one or more categories can be assigned to a document [8]. For Example, Document on Apple's iPod, may be relevant to category "audio-peripheral" as well as to "MP3 player" [1].

In text categorization problem, there is a finite set of Documents = {$d_1$, d2 .......$d_n$}, where 'n' is a very large number. The process of categorization assigns each document to a predefined set of Classes or Categories C= {$C_1$, $C_2$.....$C_k$}. Two main approaches of Knowledge Engineering and Machine Learning are used to develop various text categorization techniques. In *knowledge engineering* the classifier is tuned on a set of rules defined by knowledge engineers or domain experts. Whereas, in *Machine Learning* classifier is built by observing characteristics of manually categorized documents in initial corpus. Machine Learning approach does not involve any human intervention, that is, no knowledge engineers or domain experts are required to tune classifier [10].

Main applications of Text Categorization are automatic indexing for Boolean information retrieval systems, document organization, text filtering, and word-sense disambiguation [6]. Automatic indexing for Boolean information retrieval systems is tagging of documents by meta-data under various features like creation date, document type, document format, etc. Document organization is defined as organizing document on the basis of controlled vocabulary, which may include noise as well, and can slow down the performance of classifier. Text filtering is described as dividing text into relevant or irrelevant types. Word-sense disambiguation is termed as identifying the meaning of an ambiguous word in a text according to its semantics.

The most conventional method for document representation is BOW approach, which represents a document as an unordered collection of words; disregarding the order of words. One way to improve text categorization is to add semantic background knowledge to documents using BOW approach. Semantic background knowledge can be retrieved from various resources like WordNet, OPD, and Wikipedia, etc. WordNet is a lexical database of English language. Elberrichi, et al. [19] use WordNet; Gabrilovich, Markovitch [3] use OPD, Gabrilovich[4][5], and Pu Wang[12][13] use Wikipedia knowledge to improve text categorization by retrieving knowledge from these resources. However, the best text categorization was done by integrating knowledge from Wikipedia [4].

In this paper, we propose a way to improve categorization by adding semantic knowledge from a knowledge base Wikitology [20], where information is extracted from Wikipedia in structured and unstructured forms. The experiments are carried out by adding knowledge from knowledge base in different combinations which are using different text representations and text enrichment techniques. To enrich document, we have added k-WikiTitles, k-Categories and k-Linked concepts. We have carried out experiments using different threshold values of k. Our Experimental results demonstrate that incorporating knowledge retrieved from Wikitology shows best results up till now. An improvement of +6.36%, +6.96% and +6.99% is shown as compared to baseline and other results of enrichment by Wikitology on datasets on 20-Newsgroup, Reuters-21578(10 categories) and Reuters-21578(90 categories), respectively. The accuracy of experimental results will be evaluated by Micro-average and Macro-average F-Measure.

The organization of this paper is as follows: Section 2 discusses the related work carried out in this domain. For this, we have discussed enrichment of documents from different knowledge repositories through which text categorization was improved. Brief introduction of Wikitology is given in Section 3, along with its comparison with most up to date encyclopedia Wikipedia. In Section 4, different ways of integrating Wikitology knowledge into text document by different text representations and text enrichment techniques have been described. Section 5 illustrates empirical evaluation of experiments carried out on datasets of 20- Newsgroup, Reuters-21578 (10 categories) and Reuters-21578 (90 categories).

## 2. Related Work

In traditional approach to text categorization, features from the text are extracted and used for performing actual categorization. Features like words, phrases, sequences and concepts are very effective in determining the class relationship for a particular document. Text document in general are not very rich in such features, which can act as a core-idea for the classification algorithm. In order to overcome this problem of limited core words in the document, various researchers suggested the use of external knowledge base to enrich the document representation by simply adding the relevant knowledge to the document representation. Researchers have used Open Directory Project- (OPD) [11], WordNet [18] and Wikipedia [16] [17] for extracting relevant knowledge and enhanced the document representation. OPD based enrichment could not able to provide good results as OPD is not very well organized, its hierarchy are extremely unbalanced hence it could not offer a balance enrichment which eventually fails to give good results. WordNet is a thesaurus for English language. Elberrichi et al. [19] proposed a model in which, for all the terms, the general concepts are extracted from WordNet. Their proposed model generates Bag-of-words (BOW) of text document to be classified. The terms in the document are mapped to concepts. Three different mapping strategies are discussed: (i) add concept, (ii) replace terms with concept and (iii) remove terms and add concepts only. In order to resolve word sense disambiguation, two strategies are used: (i) to consider all concepts, (ii) to consider first concept only. Then hypernyms are added to documents. After mapping, document profile is created. For appropriate feature selection, chi-square reduction is used. Category profiles are created by selecting the features and assigning weights to them. The distance between categories profile and profile of document to be classified is calculated by using Cosine in classification phase. The proposed model was evaluated on Reuters-21578 and 20-Newsgroup dataset. Improvements from 0.649 to 0.714 and 0.667 to 0.719 are shown on Reuters-21578 and 20Newsgroup dataset respectively. Gabrilovich and Markovitch [3] suggested a model in which feature generator is built by mapping the terms of a document to OPD. They developed feature generator because OPD has non-uniform coverage, duplicate sub-trees, and unbalanced hierarchy. Feature generator works as nearest neighbor classifier and it also performs concepts generalization. Terms are mapped to OPD nodes (concepts) and features are generated by contextual analysis of the document. It also addresses two main problems of polysemy and synonymy by word sense disambiguation, and attribute selection is done by using the information gain. Feature selection removes irrelevant features. Two experiments were setup by utilizing feature selection: (i) feature selection was only applied to generated features and chosen ones were integrated into BOW of text document, (ii) no feature selection was applied on generated features and BOW were also eliminated, that is, only generated features are evaluated. Model is evaluated on datasets of Reuters-21578, RCV1, 20-Newsgroup and Movies. Evaluated results have shown improvement on all the datasets.

Wikipedia is the world's fastest growing encyclopedia. Tremendous work has been done by amalgamating Wikipedia information in different methods, to the text document. Gabrilovich and Markovitch [4] utilize Wikipedia knowledge and put forward the model in which feature generation is done through multi-resolution approach. In other words, features are generated for each document on 4 levels: individual words, sentences, paragraphs and whole text document. It maps text fragment to Wikipedia articles; but all Wikipedia articles are not considered for mapping. Mapped Wikipedia article should not have less than 100 non-stop words, not less than 5 outgoing and incoming links, and should not describe any specific dates as well as Wikipedia disambiguation pages. The problem of polysemy is addressed by determining the correct sense of each word with the help of its neighbors. Extra features are removed by using feature selection by keeping only highly discriminative features. Through feature generator text document is classified onto Wikipedia concepts. Attribute selection is done to reduce noise, in which each attribute represents a concept. Furthermore, inverted index is built to enhance vector matching. In short, in this approach basically documents are matched with most relevant documents without considering thesaurus of the article. Proposed model is evaluated on diverse datasets (Reuters-21578, RCV1, OHSUMED, 20-Newsgroup and Movies) and shows improvement.

Wang et al. [12] focused on Wikipedia thesaurus like synonymy, polysemy, hyponymy and associative relations (hyperlinks). They discussed 3 different content based, out-link based and distance based measures. Content based measure evaluates the relatedness of two linked articles by considering terms that appear in both articles. This relatedness is calculated by cosine similarity of term frequency-inverse document frequency (TF-IDF). Out-link based measure evaluates relatedness of a hyperlink between two articles using cosine similarity. Distance based measure is the shortest path between two conceptual nodes in acyclic graph formed by Wikipedia hierarchical categorization structure. Linear combination measure is constructed by combining content-based, out-link based and distance based measure. Plus, linear combination measure is adjusted on some parameters after running experiments. Text document enrichment is done by indexing Wikipedia concepts, then searching Wikipedia concepts in documents and adding Wikipedia concepts into document. Disambiguation resolution is addressed at the level of text similarity and context of document. Text document is enriched by integrating Wikipedia thesaurus and it is evaluated on Reuters-21578 and 20-Newsgroup datasets. This approach shows more improvement than the one proposed in [4].

Gabrilovich and Markovitch [5] proposed a novel method of semantic relatedness which is calculated using Wikipedia-based explicit semantic analysis, in which concepts obtained from Wikipedia are represented in high-dimensional space. Semantic interpreter is constructed which builds weighted inverted index of Wikipedia concepts related to input text fragment. Inverted index is built to remove inconsequential relations between words and concepts and this is done by observing the concepts whose TF-IDF weight is too low for a particular given word. Semantic interpreter is implemented as a centroid based classifier, which ranks Wikipedia concepts by their relevance to text fragment. Next, using semantic interpreter, weighed vector of Wikipedia concepts is built, followed by vector comparison. Relatedness of vectors is computed by cosine metric. Improvement is shown in computing word or text relatedness.

## 2.1 Wikitology

### 2.1.1 Comparing Wikipedia and Wikitology

Wikipedia is a fast-growing knowledge repository. It publishes various types of meta-data and across its pages are many thousands of micro-formats. It does not place cookies as well. It is a well-organized knowledge base which enables and provides accessibility and inter-linking of structured, semi-structured, and unstructured information. It enhances searching and browsing by interlinking of categories and articles. Each Wikipedia article is represented as a concept. It has a non-hierarchical structure. Through page linking, Wikipedia helps to establish meaningful topic associations between different pages. Hyperlinks between articles have many semantic relations such as equivalence relation (Synonymy), hierarchical relations (hyponymy) and associative relations [12]. As Wikipedia is an open-source, it contains much noise.

Wikitology is a hybrid knowledge base of information (structured and unstructured) extracted from Wikipedia. It uses a specialized information retrieval index comprised of text, instance fields, and reference fields. Knowledge in Wikitology is represented in different data structures, like: an IR index, graphs (category links, page links and entity links), relational database and a triple store. Applications can access information in knowledge base by using either simple free text queries or complex queries over multiple index fields. Structured data has been produced by sources like DBpedia and Freebase. Through reference field, applications can process information in any specific ontology by using YAGO ontology [20]. Figure 1 illustrates the Wikitology indices.

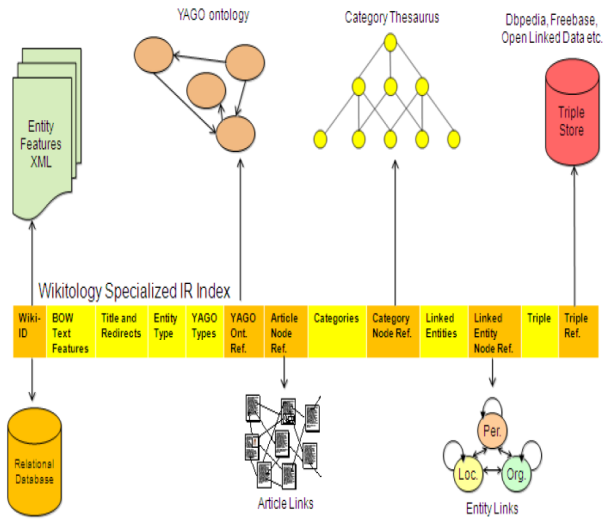

**Figure 1: Wikitology specialized information retrieval index [20]**

## 2.2 Wikitology IR Indexes

In this section, we briefly describe the information retrieval indexes of Wikitology used by us to improve content-based text categorization.

### 2.2.1 Bag of Word (BOW) Text Features

This field contains bag of words text features extracted from Wikipedia articles [20].

### 2.2.2 Titles and Redirects

Title is defined as topic of a Wikipedia article. In Wikipedia, each article represents a concept. Redirects are defined as alternative name of the concept which may include spelling deviation, synonyms, abbreviations, colloquialisms, and scientific terms [12]. This field contains titles of concepts and redirects to those concepts in Wikipedia. [20] For Example, there are 111 redirects for a concept "Barack Obama", which comprises of matches like Bacak_Obama, President_Barack_Hussain_Obama, Senator_Barack_Obama, Barack_O'Bama etc.

### 2.2.3 Entity Types

Entity is defined as a set of terms (generally nouns) that share same properties or attributes. This field contains labeled Wikipedia concepts by Freebase resource as entity types, such as Person, Location or Organization [20]. For example, we know that "Barack Obama" is the president of United States of America that is, he is a person. So, in this field of entity types, this concept will be marked as "Freebase:person".

### 2.2.4 Categories

Category is defined as general word which relates the term in is-a-kind of fashion. This field contains a list of related categories with Wikipedia article [20]. For example: categories of concept "Barack Obama" are United_States_presidential_candidates_2008, United_Church_of_Christ_members, Kenyan-Americans, University_of_Chicago_faculty, Politicians_from_Chicago etc.

### 2.2.5 Linked Entities

This field contains lists of linked Persons, Locations, and Organizations with the concept [20]. For example, for concept "Barack_Obama" the linked entities are Jeremiah_Wright, Latin_honors, Executive_Office_of_the_President_of_the_United_States, University_of_Chicago_Law_School, John_Kerry, Michelle_Obama, etc

**Table 1: Titles, Redirects, Categories, Entity Types and Linked Concepts added into 20-Newsgroup documents # 178,929 of "talk.politics.misc"**

| Concepts | Titles and Redirects | Entity Types | Categories | Linked Concepts |
|---|---|---|---|---|
| Health_insurance_in_the_United_States | Health_insurance_in_the_United_States Health_insurance_in_US Health_insurance_reform | None | Health_insurance_in_the_United_States Health_insurance Medicare_and_Medicaid_(United_States) Healthcare_in_the_United_States | American_Enterprise_Institute Congressional_Budget_Office Newborns'_and_Mothers'_Health_Protection_Act American_College_of_Physicians United_States_Census_Bureau TRICARE Medicare_(United_States) |

| Concepts | Titles and Redirects | Entity Types | Categories | Linked Concepts |
|---|---|---|---|---|
| Kaiser_Permanente | Kaiser_Permanente Kaiser_Foundation_Research_Institute Kaiser_Permanente Kaiser_Permanente_Hospital Kaiser_Permanente_entities Kaiser_Permanente_hospital | Freebase:organization | Health_care_companies_of_the_United_States Hospital_networks Non-profit_organizations_based_in_the_United_States Health_maintenance_organizations Medical_and_health_organizations_based_in_the_United_States Companies_based_in_Oakland,_California | AFL-CIO Elk_City,_Oklahoma Ohio Georgia_(U.S._state) Preventive_medicine Los_Angeles_Times Henry_J._Kaiser Centers_for_Disease_Control_and_Prevention Sicko |

## 3. Integrating Wikitology Knowledge into Text Document Representation

To improve text categorization, we will integrate semantic background knowledge retrieved from Wikitology into the text document. In this section, first we will discuss the different ways to query Wikitology knowledge base; next, we will discuss the text document representation, and finally how we have enriched the text document. Figure 2 is a diagrammatic representation of the approach that has been proposed in this paper.

## 3.1 Text Document Representation

The text document is represented as BOW. First, the document is split on delimiters into terms. For example, Original Post of 20-Newsgroup document 179,112 of "talk.politics.misc" category is

"*Why? He, Reno, and the FBI got what they wanted -- a reminder of who is the boss in America -- the thugs who work for the government.-- Clayton E. Cramer {uunet,pyramid}!optilink!cramer My opinions, all mine!*"

### 3.1.1 Different Ways of Text Representation

Following are some ways of representing text document before its enrichment by information retrieved from knowledge base.

**T1** - Stop words are removed from terms of document using stop word list. Stop words are defined as frequently used non-descriptive words. For Example:

"*reno fbi got wanted reminder of who boss america thugs work government clayton cramer uunet pyramid optilink cramer opinions*".

**T2** - Terms are tagged as entity types (Person, Location, Organization) using Stanford Named Entity Recognizer (NER)[1]. For Example

"*Why? He, <PERSON>Reno</PERSON>, and the <ORGANIZATION>FBI</ORGANIZATION> got what they wanted -- a reminder ofwho is the boss in <LOCATION>America</LOCATION> -- the thugs who work for the government.-- <PERSON>Clayton E. Cramer</PERSON> {uunet,pyramid}!optilink!cramer My opinions, all mine!*"

**T3** - After removing stop words (T1), we removed all the other words from the document which were not Noun, that is, document is left with Nouns only. This task has been achieved by using Stanford Part of Speech (POS) Tagger[2]. For example:

"*Reno FBI reminder ofwho boss America thugs government Clayton Cramer uunet cramer opinions mine*"

**T4** - After applying above method (T3) and tagging terms as entity types (Person, Location, Organization) we used Stanford Named Entity Recognizer (NER). For example:

"*Reno <ORGANIZATION>FBI</ORGANIZATION> reminder of who boss <LOCATION>America</LOCATION> thugs government <PERSON>Clayton Cramer</PERSON> uunet cramer opinions mine*"

## 3.2 Text Document Enrichment

Text Document is enriched by integrating knowledge of knowledge base. Knowledge from Wikitology is retrieved by querying the knowledge base in different ways. For more clarification, we have picked a document number 178,929 of 20-Newsgroup dataset of "talk.politics.misc" category. Selected document is of text document representation (T1) and we used different enrichment technique defined below. Table 1 illustrates the example concepts in which redirects, entity types, categories, and linked entities are retrieved from Wikitology for concepts.

### 3.2.1 Different Ways of Text Enrichment

Following are some different ways to retrieve information from Wikitology.

**E1** - Uses document representation of Type 1 or 3 (see Section 4.1). Get top N similar articles by matching the contents of the document. Then get the concepts of the titles of hit articles. Retrieve categories of the concepts. Add titles and their related categories to the document.

**E2** - Uses document representation of Type 1 or 3 (See Section 4.1). We query knowledge base using Lucene. Lucene is a full-featured text search engine library, it can search over large number of applications like searchable email, online documentation search, searchable web pages, website search, content search, version control and content management, news feed etc. We have used Lucene to search over Wikitology index file by defining search criteria on the contents of the document (which are basically terms), title of the document and also apply constraint that the page rank, should be greater than 5. We integrated wiki titles, categories and linked concepts to the document. Below is an example of (Reuters-21578 document# 6128) Lucene query use over searching knowledge base.

*Original Post: sterling drug said submitted new drug application food drug administration permission market oral form corotrope (milrinone) drug treating chronic congestive heart failure. sterling said application includes series studies 952 patients results multicenter studies involving 571 patients demonstrate efficacy safety drug alternative digitalis.*

*Lucene Query = wikiTitle:usa contents:sterling contents:drug contents:said contents:submitted contents:new contents:drug contents:application contents:food contents:drug contents:administration*

---
[1] Stanford Named Entity Recognizer (NER) labels sequences of words in a text which are the names of things in 3 classes: Person, Organization, Location. For more details see http://nlp.stanford.edu/software/CRF-NER.shtml

[2] Stanford Part of Speech Tagger assigns parts of speech to each word (and other token), such as noun, verb, adjective, etc. For more details see http://nlp.stanford.edu/software/tagger.shtml

*contents:permission contents:market contents:oral contents:form contents:corotrope contents:(milrinone) contents:drug contents:treating contents:chronic contents:congestive contents:heart contents:failure. contents:sterling contents:said contents:application contents:includes contents:series contents:studies contents:952 contents:patients contents:results contents:multicenter contents:studies contents:involving contents:571 contents:patients contents:demonstrate contents:efficacy contents:safety contents:drug contents:alternative contents:digitalis. -pageRank:[1 TO 5]*

**E3** - Uses document representation of Type 2 or 4 (See Section 4.1), in which terms in a document are already tagged in entity types: Person, Location and Organization. Above query (2) is improved by setting the types reference to Freebase Person, Location or Organization.

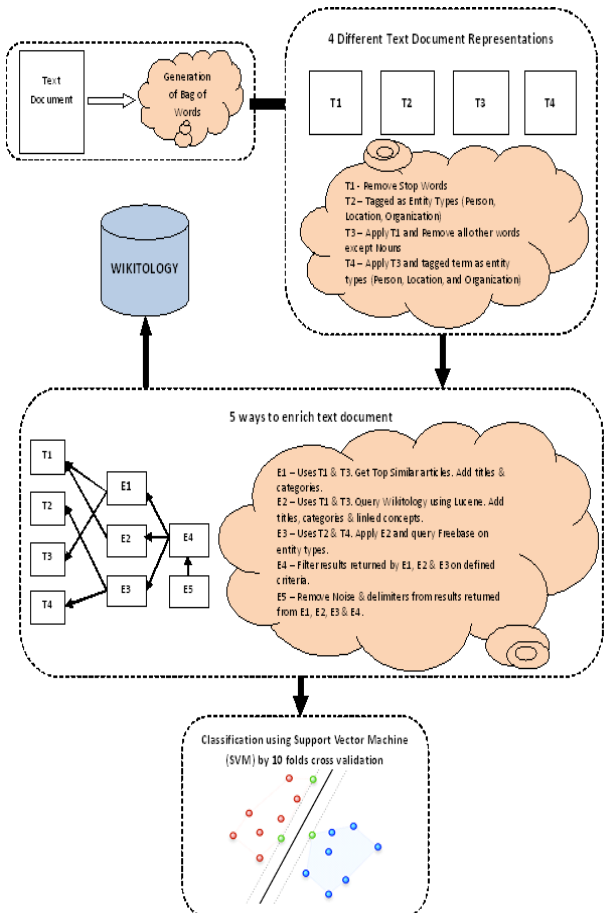

**Figure 2: Suggested Approach.**

**E4** – In this technique we have to check and filter the results (that is hits by Wikitology) returned by above three queries (E1, E2 and E3). The checking or filtering criteria of results is as follows:

(a) Result's first character should be capital, that is, in upper case and
(b) It should not contain any number

The idea of filtering results returned by Wikitology is inspired by Pu Wang et al. [12], where before indexing Wikipedia articles they remove useless and improper titles belonging to chronology (years, decades, centuries) and titles with first letters in lower case.

**E5** - To remove all noise (i.e. stop words) and delimiters from results returned by above four ways (E1, E2, E3 and E4).

## 4. Experimental Setup

### 4.1 Dataset

We used two standard document datasets to compare the quality of our proposed approach. These data sets are selected mainly due to the fact that most researchers whom work is related to this study have used the same datasets to report their results and comparisons. These datasets are:

1. Reuters: The Reuters-21578[14], test collection of Distribution 1.0 is used. The collection appeared in Reuter's newswire in the year 1987. The collection consists of 22 data files, an SGML DTD file describing the format of the available data, and six files describing the categories used to index data. The collection is available at http://www.daviddlewis.com/resources/testcollections/reuters21578/

In the Reuters-21578, we used the ModApte split (9603 training and 3299 testing documents), and two category sets: the 10 largest categories, and 90 categories with at least one training example and one testing example, similar to [14].

2. NEWS20: NEWS Group [9] is also a popular data set among text mining community; it's mainly used for text classification and clustering measure for machine learning techniques. The data set consists of approximately 20,000 newsgroup documents, partitioned in 20 different classes. The data set is available at http://people.csail.mit.edu/jrennie/20Newsgroups/

### 4.2 Evaluation metrics

We evaluated the categorization task with F-measure, which is a harmonic mean to recall and precision. We have used two types of F-measures, (i) Micro-averaged F-Measure: In micro-averaging, F-measure is computed globally over all category decisions. Micro-averaged F-measure gives equal weight to each document and is therefore considered as an average over all the document/category pairs. It tends to be dominated by the classifier's performance on common categories. (ii) Macro-averaged F-Measure. In macro-averaging, F-measure is computed locally over each category first and then the average over all categories is taken. Macro-averaged F-measure gives equal weight to each category, regardless of its frequency. It is influenced more by the classifier's performance on rare categories.

### 4.3 Experimental Results

The datasets are preprocessed and stop words and delimiters are removed. The Porter Stemmer algorithm is applied later on the preprocessed data. We have set a baseline for our experiment by simply using the data after these two steps of preprocessing. A linear Support Vector Machine-SVM [2] [15] is used to learn a model of classification. We have used micro-averaged and macro-averaged F-measure for evaluation of experimental studies; since the dataset's categories are substantially differ in sizes. In order to validate our classification data we used 4 fold cross- validations and used paired t test to assess the significance.

#### 4.3.1 The Effect of Document Enrichment

We have performed intensive experiments by combining both, text document representation scheme suggested in this paper (Sec. 4.1) and text document enrichment schemes (Sec. 4.2) techniques. Baseline shows the result of the experiment without adding semantic background knowledge. In Table 2, rows, A1 and A2 show the performance of datasets by augmenting knowledge (top 5 and 20 categories and titles) using combined enrichment techniques E1, E4 and E5 on text document representation T1. Here A1 denotes top 5 titles and their related categories and A2 denotes top 20 titles and their related categories. A3 and A4 show the performance of datasets by integrating knowledge (top 5 and 20 categories, titles and linked concepts) using combined enrichment techniques E2, E4, and E5 on text document representation T1. A3 denotes top 5 titles, their related categories, and linked concepts; A4 denotes top 20 titles, their related categories and linked concepts. A5 shows the performance of dataset by incorporating data (top categories, titles and linked entities) using combined enrichment techniques E1, E2, E4 and E5 on text document representation T1. A5 denotes top 20 titles, their related categories, and linked concepts.

Table 2, also illustrates the evaluation results of experiments (A1, A2, A3, A4, A5) carried out on datasets of 20-Newsgroup, Reuters-21578 (10 categories) and Reuters-21578 (90 categories). This table also shows the comparison of results of integration of Wikipedia knowledge by Gabrilovich and Markovitch [5] with our experimental results, that is, result achieved by integrating Wikitology knowledge in combinations of 4 text document representation and 5 text enrichment techniques. Wikipedia results from [5], have been compared with A4 technique. In our experiment A4 is the best representation as we get high micro and macro F-score on baseline data with this representation. The result of the experiments clearly established the difference between Wikipedia and Wikitology. By augmenting Wikipedia knowledge using 20-Newsgroup, Reuters-21578 (10 categories) and Reuters-21578 (90 categories) datasets an improvement of +1.0%, +1.5% and +0.7%, respectively was achieved by Gabrilovich and Markovitch [5]. In contrast, by incorporating Wikitology knowledge using 20-Newsgroup, Reuters-21578 (10 categories) and Reuters-21578 (90 categories) datasets improvement of +6.36%, +6.96% and +6.99%, respectively was observed. A comparison in Table 2 show that addition of wikitology information is producing far better results in contrast to Wikipedia.

**Table 2: Result of the experiments**

| | Wikipedia | | Wikitology | | Improvement by Wikipedia | | Improvement by Wikitology | |
|---|---|---|---|---|---|---|---|---|
| | Micro | Macro | Micro | Macro | Micro | Macro | Micro | Macro |
| **20-Newsgroup** | | | | | | | | |
| Baseline | 0.854 | - | 0.868 | 0.865 | - | - | - | - |
| A1 | - | - | 0.784 | 0.768 | - | - | -9.68% | -11.21% |
| A2 | - | - | 0.770 | 0.757 | - | - | -11.29% | -12.49% |
| A3 | - | - | 0.843 | 0.830 | - | - | -2.88% | -4.05% |
| A4 | 0.862 | - | 0.919 | 0.920 | +1.0% | - | +5.88% | +6.36% |
| A5 | - | - | 0.851 | 0.839 | - | - | -1.96% | -3.01% |
| **Reuters-21578 (10 Categories)** | | | | | | | | |
| Baseline | 0.925 | 0.874 | 0.930 | 0.905 | - | - | - | - |
| A1 | - | - | 0.854 | 0.832 | - | - | -8.17% | -8.06% |
| A2 | - | - | 0.847 | 0.802 | - | - | -8.92% | -11.38% |
| A3 | - | - | 0.898 | 0.872 | - | - | -3.44% | -3.64% |
| A4 | 0.932 | 0.887 | 0.955 | 0.968 | +0.8% | +1.5% | +2.68% | +6.96% |
| A5 | - | - | 0.937 | 0.925 | - | - | +0.93% | +2.20% |
| **Reuters-21578 (90 Categories)** | | | | | | | | |
| Baseline | 0.877 | 0.602 | 0.865 | 0.643 | - | - | - | - |
| A1 | - | - | 0.790 | 0.632 | - | - | -8.67% | -1.71% |
| A2 | - | - | 0.756 | 0.616 | - | - | -12.60% | -4.19% |
| A3 | - | - | 0.836 | 0.640 | - | - | -3.35% | -0.46% |
| A4 | 0.883 | 0.603 | 0.909 | 0.688 | +0.7% | +0.2% | +5.08% | +6.99% |
| A5 | - | - | 0.879 | 0.677 | - | - | +1.61% | +5.28% |

The best result are achieved by adding semantic background knowledge to enhance text categorization for 20-Newsgroup dataset is of 0.919 (micro-average F-Measure) and 0.920 (macro-average F-Measure) with improvement of +5.88% and +6.36%, respectively, as compared to simple baseline. Similarly, for Reuters-21578 (10 cat.) best improvement is of 0.955 (micro-average F-Measure) and 0.968 (macro-average F-Measure) with improvement of +2.68% and +6.96% respectively. This improvement for all datasets is achieved in experiment A4. We also investigate the poor performance of A2, for 20-Newsgroup dataset with 0.770 (micro-average F-Measure) and 0.757 (macro-average F-Measure) with decline of -11.29% and -12.49% respectively. We identify that the reason for this poor performance is that the representation of A2 is only top 20 titles that in fact add some irrelevant knowledge thus decreases the classification accuracy. Similarly, on other dataset of Reuters-21578 under (10 categories) the micro-averaged F-measure was 0.847, and for (90 categories) it was 0.756. The results on these two clearly shows that increasing the number of classes reduce the classification accuracy. This observation is consistent with macro-averaged F-measure as well.

The experiment clearly shows that A4 representation outperforms with apparent improvement on all previous approach to document representation with enrichment. A4 shows clear improvement because of addition of top 20 titles, categories and their linked concepts. Improvement is due to filteration criteria (E5) which eliminates titles with numbers and first letter as small; which was a cause of removal of noise and irrelevant information from the document. While A2 representation performed poorly as it does not apply filteration on the top-titles. Thus noise and irrelevant to context of the document causes the decline in F-measure.

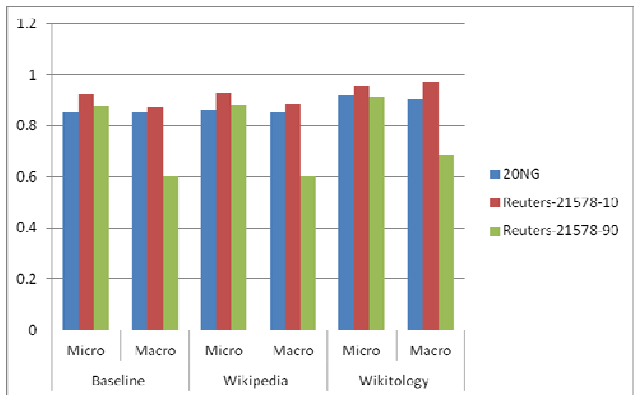

**Figure 3: Comparison of improvement achieved by Wikitology over Wikipedia**

## 5. Conclusion & Future work

In this paper, we proposed a way to integrate the semantic background knowledge retrieved from Wikitology to improve content-based text categorization. Wikitology is a knowledge base which extracts knowledge from Wikipedia and wrapped it by ontological structures. A user can query structured and unstructured information in different forms by using simple or complex queries over multiple fields. We have discussed four different ways (T1-T4) of text document representation and five ways of text enrichment (E1-E5). After carrying out extensive experiments by combining both text representations and enrichment techniques on 20-Newsgroup, we achieved an improvement of +6.0% on both micro and macro version of F-measure. Similarly, on Reuters 21578, we selected a subset of documents to form Reuters-21578 (10 categories) and Reuters-21578 (90 categories), the improvement of +6.96% and +6.99% is achieved by integrating information extracted from Wikitology. The results are clearly outperforming Wikipedia based knowledge enrichment as suggested by [4]. The improvement comes from the ontological based knowledge structures are more relevant and semantic rich hence they help in reducing the categorization errors. We believe that the Wikitology information retrieval Indexes with different structures and construct can still under utilization, it include a lot semantic rich knowledge. There are various possibilities for future extension of this work. We will try to improve text enrichment technique by applying Porter Stemmer algorithm on complete dataset after integrating semantic knowledge. Moreover, we can explore other document representation and text enrichment techniques. We can also improve the way to query Wikitology by querying information in it on more fields. We can also explore Category graph, Page links graph, and Entity links graph represented in Wikitology. Wikitology uses spreading activation algorithm to select most appropriate terms by aggregating and consolidating results of the search. Category graph can be used to predict generalized concepts while article link graphs help by predicting more specific concepts and concepts not present in category hierarchy. Category and article link graphs are used to predict concepts common to a set of documents. Page link graphs help to suggest new category concepts, identified as union of pages [20]. We can also determine a distance measure for these graphs by tuning some parameters. The improvement in text categorization under document enrichment through Wikitology is clearly due to the multiple semantic features that can easily be extracted and incorporated into the categorization approaches.

**Acknowledgments**

We would like to acknowledge the anonyms reviewers for their valuable comments on the submitted paper. We would also like to thanks National University of Computer

& Emerging Sciences, for its kind support in carrying out this research project.

**Muhammad Rafi** received his MS degree in computer science with a Gold Madel from National University of Computer & Emerging Science, FAST-NU Karachi campus in 2000. He is currently pursuing his PhD degree in Computer Science from the same university. He is an assistant professor in computer science department of FAST-NU, Karachi Campus. His research interests include machine learning, algorithm design, data/text mining and information retrieval. He is also a member of ACM and IEEE.

**Sundus Hassan** received her MS in computer science in 2010. She is currently a software engineer at a local software company.

**Muhammad Shahid Shaikh** received the BE degree from Mehran University of Engineering and Technology, Pakistan, in 1986, the MS from Michigan State Univeristy in 1989 and the PhD degree from McGill University, Montreal, in 2004, all in Electrical Engineering. He is currently associate professor and head of the department of Electrical Engineering at the National University of Computer and Emerging Sciences, Karachi, Pakistan.